\documentclass{amsart}
\DeclareMathOperator{\vol}{vol}

\DeclareMathOperator{\IM}{Im}

\DeclareMathOperator{\ad}{ad}

\theoremstyle{plain}
\newtheorem{theorem}{Theorem}[section]

\newtheorem{proposition}[theorem]{Proposition}

\newtheorem{corollary}[theorem]{Corollary}
\newtheorem{conjecture}[theorem]{Conjecture}

\theoremstyle{definition}

\theoremstyle{remark}

\numberwithin{equation}{section}
\numberwithin{figure}{section}

\newcommand{\cH}{{\mathcal H}}

\newcommand{\cA}{{\mathcal A}}
\newcommand{\cE}{{\mathcal E}}
\newcommand{\cK}{{\mathcal K}}

\newcommand{\cD}{{\mathcal D}}

\newcommand{\cM}{{\mathcal M}}
\newcommand{\cC}{{\mathcal C}}
\newcommand{\cG}{{\mathcal G}}
\newcommand{\cR}{{\mathcal R}}
\newcommand{\tR}{\widetilde{\cR}}
\newcommand{\tk}{\tilde{k}}

\newcommand{\cW}{{\mathcal W}}

\newcommand{\fc}{\mathfrak c}
\newcommand{\fp}{\mathfrak p}
\newcommand{\fg}{\mathfrak g}

\newcommand{\ft}{\mathfrak t}

\newcommand{\CC}{{\mathbb C}}
\newcommand{\HH}{{\mathbb H}}
\newcommand{\RR}{{\mathbb R}}
\newcommand{\ZZ}{{\mathbb Z}}

\newcommand{\rd}{\mathrm{d}}

\newcommand{\<}{\langle}
\renewcommand{\>}{\rangle}

\begin{document}

\title[Moduli space of monopoles]{A note on  monopole moduli spaces}
\author{Michael K. Murray}
\address[Michael K. Murray]
{Department of Pure Mathematics\\
University of Adelaide\\
Adelaide, SA 5005 \\
Australia}
\email[Michael K. Murray]{mmurray@maths.adelaide.edu.au}
\author{Michael A. Singer }
\address[Michael A. Singer]
{Department of Mathematics and Statistics \\
James Clerk Maxwell
Building \\
University of Edinburgh EH9 3JZ \\
U.K.}
\email[Michael A. Singer]{michael@maths.ed.ac.uk}


\begin{abstract}
  We discuss  the structure of the framed moduli space of Bogomolny
monopoles for arbitrary symmetry
breaking and extend the definition of its stratification to the case
of arbitrary compact Lie groups. We show that each stratum is a union
of submanifolds for which we conjecture that the natural $L^2$ metric
is hyperK\"ahler. The dimensions of the strata and of these submanifolds
are calculated, and it is found that for the latter, the dimension is
always a multiple of four.
\end{abstract}
\maketitle

\section{Introduction}
Recently there has been much interest in monopoles with non-maximal
symmetry breaking at infinity. In particular questions have been raised
as to when they are manifolds and when they have hyperK\"ahler metrics.
This note gathers together some mathematical
results concerning the structure of the moduli spaces and their $L^2$
metrics.  These range
from theorems which have been proved in full generality, through partially
proved theorems to outright conjectures.

Recall that we  generally expect that moduli spaces of solutions of
the self-duality equations  and their
reductions such as the Bogomolny equations and Nahm's equations
should be
hyperK\"ahler manifolds. One reason for this is that formally such
moduli spaces arise as
hyperK\"ahler quotients. To recall this, fix a compact, connected Lie
group $G$, with Lie algebra $\fg$, and consider the space $\cA$ of
$G$-connections (vector potentials) on the trivial $G$-bundle over
flat $\RR^4$. By identifying
$$ A_0dx_0 +  A_1dx_1 +  A_2dx_2 + A_3dx_3
$$
with the $\fg\otimes \HH$-valued function
$$
A_0 + i A_1 + j A_2 + k A_3,
$$
where $i$, $j$ and $k$ are unit quaternions, $\cA$
becomes a quaternionic vector space. Formally,
$\cA$ can be equipped with the $L^2$-metric, making it a flat
hyperK\"ahler manifold.  Because $\RR^4$ is not compact, the
convergence of this metric will depend upon subjecting our connections
to suitable asymptotic conditions, and these will be considered in
detail below.  Setting this aside for the moment, it is a
straightforward exercise to
check that the hyperK\"ahler moment map for the
action of the gauge group $\cG$ on $\cA$ is given by
$$
A\longmapsto F^+_A\in
\Omega^2(X,\fg)\otimes \IM \HH.
$$
Hence the hyperK\"ahler quotient $\cA/\!/\!/\!\cG$ should be the same
as the space of anti-self-dual connections divided 
by the action of the gauge group, and the
$L^2$ metric will descend to define a hyperK\"ahler metric
on the moduli space.

A monopole on $\RR^3$ is a pair $c = (A,\Phi)$, where $A$ is a
connection on the trivial $G$-bundle $E\to \RR^3$, and $\Phi$ is a
section of the adjoint bundle $E\times_G \fg$. The monopole $c$
satisfies the Bogomolny equations
\begin{equation}\label{bogo1}
\rd_A \Phi = * F_A
\end{equation}
if and only if the connection $\Phi dx_0 + A$ is anti-self-dual on
$\RR\times \RR^3$.  In particular, from this four-dimensional point of
view, $\Phi$ cannot vanish at infinity, because it is independent of
$x_0$. Thus the convergence of the $L^2$ metric and the non-degeneracy
of the hyperK\"ahler symplectic forms are important issues in this case.

These issues were fully resolved when $G =SU(2)$ by Atiyah and Hitchin
\cite{AtiHit}: they showed  that the moduli space
of (framed) monopoles of charge $k$ is, indeed, a complete hyperK\"ahler
manifold. Its dimension is $4k$ where the charge of the monopole is
$k$.

For a general compact Lie group of rank $r$ it is expected that the
moduli space of monopoles with {\em maximal symmetry} breaking is a
hyperK\"ahler manifold although this has not been proved in
generality. Except for very simple low charge cases,  there 
are mostly partial results which
compute the metric asymptotically near the edge of the moduli space,
see for example \cite{Bie1, Bie2, LeeWeiYi96} and references therein.

The real complications however arise when there
is {\em non-maximal symmetry breaking} which is our primary interest
below. The case of $SU(3)$ monopoles with minimal symmetry
breaking was treated in detail by  \cite{Dan},   but beyond this
little seems to be known.

\vspace{10pt}
We shall present here a summary of the results discussed in
the paper: the reader will have to refer forward for precise definitions.

The full moduli space of (framed) monopoles of mass $\mu$ and charge
$m$ is denoted by $\cM(u,\mu,[\phi]=m)$. Here $0\not=\mu\in \fg$ is
arbitrary\footnote{{\em maximal} symmetry breaking is precisely the
condition that $\mu$ should be regular}
and $u$ is a unit vector in $\RR^3$. $m$ is a homotopy
class, essentially a string of integers. The boundary conditions
imposed guarantee that for some $k\in \fg$,
\begin{equation}\label{3.10.12}
\Phi(tu) = \mu - \frac{k}{2t} + o(t^{-1}) \mbox{ for }t\gg 0.
\end{equation}
There is therefore a map $e:\cM(u,\mu,[\phi]=m) \to \fg$ which assigns
$k$ to $(A,\Phi)$. The image $\cK$ of $e$ in $\fg$ is not the whole of
$\fg$, but rather a disjoint union of $C(\mu)$-orbits
\begin{equation}\label{4.10.12}
\cK = C(\mu)k_1\cup C(\mu)k_2\cup\ldots\cup C(\mu)k_n.
\end{equation}
It turns out that the $k_j$ are {\em integral} elements of $\fg$. The
set of all monopoles $(A,\Phi)$ with $e(A,\Phi)\in C(\mu)k_j$ is the
$j$-th stratum $\cM_j$, say, of the moduli space. This was defined in
a different
way for $G = SU(r+1)$ in \cite{Mur}. In general, $\cM_j$ does not have
dimension divisible by $4$, so cannot be hyperK\"ahler. However, if we
define, for $k\in \cK$,
\begin{equation}\label{5.10.12}
\cM(u,\mu,k) = \{(A,\Phi)\in \cM(u,\mu,[\phi]=m): e(A,\Phi) = k\}
\end{equation}
(the moduli space of framed monopoles of {\em type} $(\mu,k)$) then we
shall see that $\cM(u,\mu,k)$ has dimension divisible by $4$ and the
natural conjecture is that the $L^2$ metric makes $\cM(u,\mu,k)$ into
a hyperK\"ahler manifold.

At least one of the strata, $\cM_1$, say, must be open, hence of the
same dimension as $\cM(u,\mu,[\phi]=m)$, but this stratum need not be
hyperK\"ahler. If, however, $C(\mu)k_1 = k_1$ then $\cM_1 =
\cM(u,\mu,  k_1)$ and
then this stratum {\em is} a
candidate to be hyperK\"ahler.  Notice more generally that if $k$ and
$k'$ lie in $C(\mu)k_j$, an element $g\in C(\mu)$ with $\ad(g)k = k'$
can be regarded as a constant gauge transformation which maps
$\cM(u,\mu,k)$ diffeomorphically to $\cM(u,\mu,k')$.

In \S\ref{sec:charges}, magnetic charges $m_1,\ldots, m_s$ and holomorphic
charges $h_1,\ldots, h_{r-s}$ are defined for monopoles in $\cM(u,\mu, k)$.
The information in the magnetic charges is topological and is
equivalent to the homotopy class $m$. In particular, the magnetic charges
do not vary from stratum to stratum. By
contrast the holomorphic charges determine the stratum $\cM_j$. (The
number $s$ of magnetic charges is completely determined by the mass
$\mu$.)

We shall show
that if $\cM(u,\mu,k)$ is non-empty, then the charges are all
non-negative, and that
\begin{equation}\label{7.10.12}
\dim \cM(u,\mu,k) = 4(m_1+ \ldots + m_s + h_1 + \ldots + h_{r-s}).
\end{equation}
Dimensions of the strata and full moduli space are also determined in
\S\ref{sec:dim}.

\section{The moduli space as a manifold}

In this section we shall introduce various different monopole moduli
spaces and explain carefully which of them are smooth manifolds, and which are
likely to admit hyperK\"ahler metrics. Throughout we shall be
considering {\em euclidean} monopoles, that is to say monopoles on flat
$\RR^3$. Note that the metric enters the Bogomolny equation
\eqref{bogo1} through the Hodge star operator. Some work has also been
done on hyperbolic monopoles, where $\RR^3$ is replaced by hyperbolic
$3$-space $\cH^3$.  It is expected that moduli spaces of
hyperbolic monopoles will be diffeomorphic to  the corresponding
moduli spaces of euclidean monopoles, but this has not been proved in
general. On the other hand, the issue of natural metrics on moduli
spaces of hyperbolic monopoles is completely open: all that is known
for certain is that the $L^2$ metric is infinite in this case.

There are two reasons why there are so many different
monopole moduli spaces. The first is that the monopoles must be {\em
framed}, and this can be done either at a base-point in $\RR^3$ or `at
infinity'. The second has to do with the specification of the
asymptotics of the Higgs field $\Phi$.

\subsection{Notation}
\label{notation} In order to discuss monopoles, we shall fix the
following:
\begin{itemize}
\item $G$ is a compact, connected, semi-simple Lie group of rank
         $r$. The complexification is denoted $G^c$ and Lie algebra $\fg$.

\item if $a\in \fg$, $O_a\subset \fg$  is the orbit of $a$ in $\fg$
under the adjoint action
         of $G$. $C(a)\subset G$ is the centraliser of $a$,       with
         Lie algebra $\fc(a)$.

\item As a homogeneous space, $O_a = G/C(a) = G^c/P_a$, where $P_a$ is
         the appropriate parabolic subgroup. The latter description
         gives $O_a$ the structure of a compact complex manifold.

\item $\mu$ and $k$ are commuting elements of $\fg$, $[\mu,k]=0$.

\item $E\to \RR^3$ will denote the trivial principal $G$-bundle over
         $\RR^3$.
\end{itemize}

\subsection{Boundary conditions and moduli spaces}

The physically natural condition to impose on solution of the
Bogomolny equations is the finite-energy condition
\begin{equation}\label{finerg}
\int |F_A|^2 = \int |\rd_A\Phi|^2 < \infty.
\end{equation}
We shall impose apparently rather stronger asymptotic conditions. It
follows from the work of Taubes if $G= SU(2)$ that \eqref{finerg}
together with \eqref{bogo1} implies these stronger conditions, but for
general groups this must remain a conjecture.

Following Jarvis, we assume:
\begin{quote}
(BC1) Along each straight line, there is a gauge in which
$$
\Phi = \mu - \frac{k}{2r} + O\left(\frac{1}{r^{1+\delta}}\right)
$$
for all sufficiently large $r$,

\noindent(BC2) In this same gauge,
$$
\rd_A\Phi = \frac{k}{2r^2}\rd r + O\left(\frac{1}{r^{2+\delta}}\right)
$$
for all sufficiently large $r$.
\end{quote}
These conditions are closely related to the
Bogomolny-Prasad-Sommerfield (BPS) boundary conditions of
\cite{HurMur}.

Define
$$\cC = \{(A,\Phi): \rd_A\Phi = *F_A,\;(A,\Phi)\mbox{ satisfies BC1
and BC2}\}.
$$
Notice that we do not yet fix $\mu$ and $k$: we merely assert that the
boundary conditions are satisfied for {\em some} elements $\mu$ and
$k$ satisfying
\begin{equation}\label{1.9.12}
\mu\not=0,\;\;[\mu,k] =0.
\end{equation}
Denote by $\cG$ the group of all automorphisms $g$  of $E$ that preserve
the boundary conditions (i.e.\ $g$ and $\nabla g$ have limits as
$r$ goes to infinity along any straight line, and the limiting values
are continuously differentiable when viewed as functions on the sphere
at infinity).  Then $\cG$ acts on $\cC$ and we would like to define
the monopole moduli space as the quotient $\cM = \cC/\cG$.  This will have
singularities because $\cG$ does not act freely. In addition it will
contain components of arbitrarily high dimension. We shall now explain
how these two problems are eliminated.

\subsection{The degree of a monopole}

The asymptotic value of $\Phi$ is a section $\phi$, say, of
$\ad(E_\infty)$, where $E_\infty$ is the restriction of $E$ to the
two-sphere at infinity. Since $\ad(E_\infty)$ is a trivial bundle, we
can view $\phi$ as a continuous map into $\fg$. By BC1, this takes
values in the adjoint orbit $O_\mu$. This orbit is preserved by the
action of gauge transformations $g$ on $E_\infty$, but $g(\phi) =
\ad(g)\phi$, so that this map is not gauge-invariant. However its
homotopy class $m = [\phi]$ {\em is} gauge invariant, because
$\pi_2(G)= 0$, so that any gauge transformation can be deformed to the
identity.  The homotopy class $m$ is called the {\em degree} of the
monopole. This discussion suggests the definition of spaces
$$
\cC(O_\mu,[\phi]=m)
$$
where the adjoint orbit as well as the homotopy class of $\phi$ are
fixed. This is referred to as the set of monopoles of mass $\mu$ and
charge $m$. Note that  $O_\mu = G/C(\mu)$.

\subsection{Radial scattering and interior framing}
Let $x\in \RR^3$ be any point. The moduli space of monopoles framed at
$x$, of mass $\mu$ and charge $m$ is the quotient
$$
\cM(x,O_\mu,[\phi]=m) = \cC(O_\mu,[\phi]=m)/\cG(x)
$$
where
$$
\cG(x) = \{g\in \cG: g(x) = 1\}.
$$
In \cite{Jar00} Jarvis proved the following:
\begin{theorem} There is a natural bijection
$$
r_x: \cM(x,O_\mu,[\phi] = m) \longrightarrow\cR(O_\mu,m)
$$
where the set on the RHS is the space of all holomorphic maps
$v:S^2\to O_\mu$, with $[v] = m$.
\label{thm1}\end{theorem}
In defining $\cR(O_\mu,m)$ recall from \S\ref{notation} that $O_\mu$
is
in a natural way a complex manifold. It is known  \cite{BoyManHurMil}
that $\cR(O_\mu,m)$ is a finite-dimensional smooth manifold, often
referred to as a space of rational maps.  It follows that our framed
moduli space can be identified with a smooth manifold. It should be
the case that $r_x$ is naturally a diffeomorphism, but to prove that one
would have to equip $\cM(x,O_\mu,[\phi] = m)$ with a smooth
structure.  Although this should be possible, we are not aware of a
detailed treatment of this issue.

\subsection{Framing at infinity and parallel scattering}

To frame  monopoles `at infinity' we pick a point $u \in S^2$, viewed
as the sphere at infinity in $\RR^3$. Returning to BC1,  we define:
$$
\cC(u,\mu, [\phi] = m) = \{(A,\Phi)\in \cC:
\lim_{t\to\infty}\Phi(tu)= \mu, [\phi] = m \}
$$
and
$$
\cC(u,\mu,k) = \{(A,\Phi)\in \cC: \Phi(tu) = \mu - k/2t+o(t^{-1})\}
$$
and introduce the corresponding gauge group
$$
\cG(u) = \{g\in \cG: \lim_{t\to\infty} g(tu) = 1\}.
$$
The corresponding moduli spaces are
$$
\cM(u,\mu,[\phi] =m) =\cC(u,\mu,[\phi]=m) /\cG(u)\mbox{ and }
\cM(\mu,k)= \cC(u,\mu,k)/\cG(u).
$$
The first of these is called the moduli space of (framed) monopoles
with mass $\mu$ and degree $m$. The second is called the moduli space
of (framed) monopoles of type $(\mu,k)$.

These can also be identified with spaces of rational maps:
\begin{theorem}
\begin{itemize}

\item[(a)] There is a natural bijection
$r_u \colon \cM(u, \mu, m) \to \tR(O_\mu, m)$.

\item[(b)] There is a natural bijection $\hat r_u \colon \cM(u, \mu, k) \to
\tR(O_{\mu k},  m)$.
\end{itemize}
\label{thm2}
\end{theorem}
Here $\tR(O_\mu, m) \subset \cR(O_\mu, m)$ is the set of {\em  based}
rational maps, that is,  those which send $u\in S^2$ to $\mu$.
In part (b),
\begin{equation}\label{orbitdef}
O_{\mu k} = G/H_{\mu k} = G^c/P_{\mu k},\mbox{ where }
H_{\mu k} = C(\mu)\cap C(k)
\end{equation}
and $P_{\mu k}$ is the corresponding parabolic subgroup.

Part (a) of this result was proved first by Donaldson
\cite{Don} for $G=SU(2)$ then
by Hurtubise \cite{Hur} for classical groups by a generalisation of
Donaldson's approach. Both parts were proved
for general $G$ by Jarvis \cite{Jar98a, Jar98b} using parallel
scattering to associate a rational map to a monopole, and nonlinear
analysis to invert this procedure.

   We note in passing that Jarvis shows that the
restriction of $r_u$ to $\cM(u, \mu, k)$ is the composition of
$\hat r_u$ with the projection $\tR(O_{\mu k}, m) \to
\tR(G^c/P,  m).$

Once again, it is not clear that smooth structures have been defined
on these framed moduli spaces. One conjectures that natural smooth
structures should exist, such that these bijections are diffeomorphisms.

As we indicated in the Introduction, it is the moduli spaces
$\cM(u,\mu,k)$ that have dimensions divisible by $4$ and which are
therefore candidates to be hyperK\"ahler spaces.
In Proposition \ref{prop:fourdim} the dimension of
$\tR(O_{\mu k}, m)$ will be explicitly computed.

\subsection{Discussion}

Let $x(t) = ut$, and consider the
bijection $r_{x(t)}$, for $t$ large, of Theorem~\ref{thm1}. It is
tempting to believe that this should approach the map $r_u$ of
Theorem~\ref{thm2}. However, they cannot be compared directly since
they have different targets. But we could  divide both sides by the
appropriate groups to get bijections
$\tilde r_u \colon \cM(O, [\phi]=m) \to \tR(O_\mu, m)/C(\mu)$ and
$\tilde r_{x(t)} \colon \cM(O_\mu, [\phi]=m) \to
\cR(O_\mu, m)/G$ and then
compare them via the natural isomorphism induced by the inclusion
of based maps into unbased maps.
    A straightforward
calculation shows that the
limit of $\tilde r_{tu}(A, \Phi)$ typically does not exist because evaluated
in co-ordinates it blows up. Some kind of
renormalization or scaling must be required to find the relationship
between the limit of $\tilde r_{x(t)}$ and $\tilde r_u$.

\section{The $L^2$ metric}

Formally, a tangent vector to $(A,\Phi)$ in  $\cC$ is a pair
$(\dot{A},\dot{\Phi})$ satisfying
the linearization at $(A,\Phi)$ of the Bogomolny equations. The $L^2$
metric gives this vector length-squared equal to
\begin{equation}\label{l2metric}
\int_{\RR^3} (|\dot{A}|^2 + |\dot{\Phi}|^2)\rd x;
\end{equation}
due to the non-compactness of $\RR^3$, this need not converge.
Looking back at BC1 and BC2, it is clear that \eqref{l2metric} cannot
converge if the variation $\dot{\Phi}$ changes $\mu$ or $k$ in
BC1.  It is natural, therefore, to focus on $\cM(u,\mu,k)$ as the
obvious candidate to carry a hyperK\"ahler metric. Our first task is
to show that if the  Bogomolny equations hold asymptotically, then the
pair $(\mu,k)$ determines the leading asymptotics of the monopole on
the whole of the two-sphere at infinity.

We begin by noting that the boundary conditions imply that the
connection $A$ restricts to give a connection $a$ on $E_\infty$ and
that BC1 gives
\begin{equation}
\Phi(tz) = \phi(z) - \frac{f(z)}{2t} + o(t^{-1})
\end{equation}
where $\phi$ and $f$ are smooth functions of $z\in S^2$ and the
framing condition is
\begin{equation}\label{1.10.12}
\phi(u) = \mu,\;\;f(u) = k.
\end{equation}
The Bogomolny equations reduce to
\begin{equation}
\label{eq:boundary_data}
\nabla f = 0, \; \nabla\phi = 0,\;F_a = \frac{f}{2}\rd\!\vol
\end{equation}
where $\rd\!\vol$ denotes the standard area-form of the unit
2-sphere. A pair $(\phi,f)$ satisfying \eqref{1.10.12} and
\eqref{eq:boundary_data} are called {\em monopole boundary data}.

We  now prove  that, up to gauge, the pair $(\phi,f)$ is completely
determined by its value $(\mu,k/2)$ at the base-point $u$.

\begin{proposition}
\label{prop:onepoint}
Let  $(\phi, f)$ and $(\phi', f')$ be   boundary data for a monopole:\\[3 pt]
(i) if $u$ and $v$ are in $S^2$ then there is a $g \in G$
such that $\phi(u) = \ad(g)(\phi(v)) $ and $f(u) = \ad(g)(f(v))$;\\[2 pt]
(ii) if there is an $h \in G$ such that $\phi(u)= \ad(h)(\phi'(u))$
and $f(u) = \ad(h)(f'(u))$ then
there is a $g \colon S^2 \to G $ such that $\phi^g = \phi'$ and $f^g = f'$.
\end{proposition}
\begin{proof}
If $\phi=0$ this is a trivial case of the results of \cite{AtiBot}
classifying equivalence classes or Yang-Mills connections over a
Riemann surface.
     We follow the proof in \cite{AtiBot}. Recall that $E_\infty \to S^2$
is a principal $G$-bundle. Then $\phi$ and $f$ can be viewed as
equivariant maps $E_\infty \to \fg$.
Fix a point $p_0 \in E_\infty$
and let $\phi(p_0) = \mu$ and $f(p_0) = k$. Because $\phi$ and $f$ are
covariantly constant they are constant along any horizontal path.
If $p \in P$ we can join $p_0$ to some point $pg$  with a
horizontal curve and then $\phi(p) = \ad(g)(\mu)$ and $f(p) = \ad(g)(k)$
as required.

    {}From the discussion in the preceding paragraph it follows that we have a map
$$
(\phi, f) \colon E_\infty \to O_{\mu k}=G/H_{\mu k},\;\; H_{\mu k} =
C(\mu)\cap C(k).
$$
Here $O_{\mu k}$ is the orbit of $(\mu, k)$.
The pre-image of the coset $H_{\mu k}$, ie the
set of all points $p$ in $E_\infty$ at which $\phi(p) = \mu$ and $f(p) = k$,
is a reduction of $E_\infty$ to $H_{\mu k}$ which we denote by $E_{\mu k}$.
If $p \in E_{\mu k}$ then
any horizontal curve is also in $E_{\mu k}$ because $\phi$ and $k$ are constant
along horizontal curves so the connection also reduces to $P_{\mu k}$.

Because $S^2$ is simply connected standard results on reduction
of bundles to their holonomy subgroups can be used  \cite{KobNom}.
It follows from the Ambrose-Singer theorem that the
holonomy subgroup at $p_0$ is the subgroup  $H \subset H_{\mu k}$
obtained by exponentiating
$k$ and that $E_{\mu k}$ reduces to a bundle $E_0$ with structure group $H$.

For the final point we need to know that $k$ is an integral element of
the Lie algebra. This is done in \cite{GodNuyOli} and in different
fashion in \cite{Jar98a}. We proceed as follows.
Because $[\mu, k] =0$ the closure of the
subgroup generated by $\exp(t\mu+sk)$ for any $t$ and $s$ will be an abelian
subgroup of $G$ so a torus and hence inside a maximal torus
containing $H$. If $\lambda$
is any weight of this maximal torus we can form an
associated line bundle which will have integer chern class $\lambda(k)$.
It follows that $k$ is an integer element of $\fg$ and that it exponentiates
to define a circle subgroup and a homomorphism $\chi \colon U(1) \to G$.

We have now reduced our original bundle to a sub-bundle $Q \to S^2$
which is a circle
bundle. It has a connection $A$ and a curvature $F$ with $*F = k/2$ a
constant so
that it is a circle bundle of degree $1$.  If $A'$ is another connection
with curvature $F' = F$ then $A - A' = a$ with $da = 0$ so $a =
d(\exp(g))$ for $g \colon S^2 \to U(1)$
and hence the connections $A$ and $A'$ are equal after a gauge transformation.

This gives us a method of constructing the original bundle,
connection and Higgs field
from the data $\mu$ and $k$.  First take the standard $U(1)$ bundle
$Q \to S^2$ with its $SU(2)$
invariant connection and fix $q_0 \in Q$ in the fibre over the point $u$.
Let $\chi \colon U(1) \to G_{\mu k} \subset G$ be the homomorphism
defined by exponentiating $k$.
We can then form $Q \times_{\chi} G$ the associated bundle using the action
$(q, k)z = (qz, \chi(z)^{-1}k)$ for $z \in U(1)$.  This inherits a connection
and the Higgs field is defined by $\hat\phi([q,k]) = \ad(k)(\mu)$.

\end{proof}

Let $\cC^\infty$ denote the set of all monopole boundary data $(\phi,f)$
and let $\cG^\infty$ be the space of all
gauge transformations at infinity, that is maps $g \colon S^2 \to
G$. Define the {\em moduli space of boundary data} to be the quotient
$\cM^\infty = \cC^\infty / \cG^\infty$.
We  have the boundary map
\begin{equation}
\label{eq:boundary_map}
\partial \colon \cM \to \cM^\infty,
\end{equation}
which sends $(A, \Phi)$ to the value of the Higgs field and
curvature at infinity.
Our reason for introducing the boundary map is that we believe that the
methods of Atiyah and Hitchin \cite{AtiHit} can be adapted to to show that
\begin{conjecture}
If $\partial(A, \Phi) = \partial(A', \Phi')$ then there is a gauge
transformation
$g$ such that $A^g - A'$ and $\Phi^g - \Phi'$ are $L^2$.
\label{l2conj}\end{conjecture}
The idea here is that if the condition holds then for some gauge
transformation $g$,  $\Phi^g$
and $\Phi'$ should agree up to order $1/r$, so that $\Phi^g- \Phi'$
will be square integrable. Similar considerations should apply to the
difference between the connections.

Let $\cG^\infty(u)$ be all gauge transformations which are the
identity at $u$ and let $\cC^\infty(u, \mu, m)$  be all pairs $(\phi, f) $
with $\phi(u) = \mu$ and $[\phi] = m$. Denote $\cM^\infty(u, \mu, m)
= \cC^\infty(u, \mu, m)/\cG^\infty(u)$.
We have the commuting diagram
$$
\begin{matrix}
\cM(u, \mu, m) & \to & \cM^\infty(u, \mu, m) \\
       \downarrow &      & \downarrow \\
\cM(O, m) & \to & \cM^\infty(O, m) \\
\end{matrix}
$$
where both vertical maps are quotienting by the group $C(\mu)$.

Conjecture~\ref{l2conj} would imply that the $L^2$ metric
is finite on each of the moduli spaces $\cM(u, \mu, k)$ of monopoles
of type $(\mu,k)$.  This suggests
\begin{conjecture}
The spaces $\cM(u, \mu, k)$ are hyperK\"ahler
manifolds.
\end{conjecture}

A natural approach to these conjectures is the analysis of the
linearization $\cD$ at $(A,\Phi)\in \cM(u,\mu,k)$
of \eqref{bogo1}. Combined with
the Coulomb gauge-fixing condition, $\cD$ becomes a coupled Dirac
operator on $\RR^3$,
$$
\cD: C^\infty(\RR^3, \HH\otimes\HH\otimes \ad(E)) \to
C^\infty(\RR^3, \HH\otimes \HH\otimes \ad(E))
$$
where $\HH$ is the regarded as the spin-bundle of
$\RR^3$. Unfortunately this operator is not `invertible at infinity'
and so it is not automatically a Fredholm operator in $L^2$. Taubes
analysed it in detail when $G= SU(2)$, but in general, rigorous
results about this operator are not available. Nonetheless, it should
be possible to find a suitable space of functions such that $\cD$
becomes a Fredholm operator, with index calculable in terms of the
type data $(\mu,k)$. Formally $\cD$ is quaternionic, so its $L^2$
null space will automatically be a quaternionic vector space with
compatible inner product.   The reader is referred to
\cite[Ch 3 and Ch 4]{AtiHit} for a detailed discussion of the case $G=
SU(2)$.

\subsection{Group actions}
Consider $\cE$, the  group of Euclidean transformations
of $\RR^3$, which  is the semi-direct product of $SO(3)$, the group 
of rotations and $\RR^3$ the
group of translations. As  the monopole bundle $E \to \RR^3$  is
trivial
the group $\cE$ acts on the connection and Higgs field, preserves
the Bogomolny equations and commutes with gauge transformations so
it acts on the full-unframed moduli space.  In general this action
disturbs the framings.  If $ x \in \RR^3$ then
the sub-group $\cE_x$ of transformations preserving $x$ which
is isomorphic to $SO(3)$, acts naturally on the moduli space of
monopoles framed at $x$.  If $u \in S^2$ then the subgroup
of $\cE_u$ of transformations preserving the line through $u$,
which is isomorphic to $SO(2) \times\RR^2$, will
act naturally on the moduli space of monopoles framed at $u$.

As well as these straightforward actions the moduli space $\cM(u,\mu,k)$ also
carries  an  action of the full group of Euclidean transformations.
For this we need a different description of this moduli space (cf.\
\cite[pp.\ 15--16]{AtiHit}).  Note
that Proposition \ref{prop:onepoint} shows that  $k$
defines a representation of the circle in $G$ hence an
associated $G$-bundle over the two-sphere. This carries a
natural $SO(3)$-action  and has a unique $SO(3)$-equivariant
connection  $a$ and Higgs field
$\phi$ such that $\phi(u) = \mu$
and $f(u) = k$.  The moduli space $\cM(\mu,k)$ is now defined to
consist of
configurations $(A, \Phi, q)$  where $(A, \phi)$ is
a monopole and $q$ is an isomorphism between
$\partial(A, \Phi)$ and $(\phi, f)$, modulo the group of gauge
transformations that approach the identity at infinity. Then
$\cM(\mu,k)$ has a natural $SO(3)$-action and can be shown to be diffeomorphic to
$\cM(u,\mu,k)$.  The subtlety is (as in the case $G= SU(2)$) that
the diffeomorphism between $\cM(\mu,k)$ and $\cM(u,\mu,k)$ is not
equivariant with respect to the copy of $SO(2)\subset SO(3)$ which
fixes the direction $u$. 

\subsection{Discussion}

Assuming that the $L^2$-metric does define a genuine hyperK\"ahler
metric on $\cM(u,\mu,k)$ there are many interesting open questions
surrounding it. First of all, there is the issue of whether it is
complete for all $\mu$ and $k$. Secondly there are questions relating
to variation of the parameters $\mu$ and $k$. It is natural to
conjecture that the metrics will vary smoothly with $\mu$ as long as
the corresponding orbit $O_\mu$ does not jump. An interesting
conjecture of Lee, Weinberg and Yi \cite{LeeWeiYi} suggests that these
hyperK\"ahler metrics should also behave well with respect to
specialization of $\mu$.  To state the conjecture, call a path
$\mu : [0,\delta] \to \fg$ a {\em regular} deformation of
$\mu_0=\mu(0)$
if $\mu(t)$ is {\em regular} for
all $t>0$. Let $\cM_t = \cM(u,\mu_t,k)$, and let $g_t$ be the $L^2$
metric on $\cM_t$.

\begin{conjecture} Given any $0\not=\mu_0\in \fg$,
there is a regular deformation $\mu_t$, such that $(\cM_t,g_t)$ tends
to $(\cM_0, g_0)$ as $t\to 0$.
\end{conjecture}

Note that Jarvis \cite{Jar98a} describes  a `filling-out procedure'
which associates to any holomorphic map  $v:S^2\to O_{\mu,k}$ a
new map $\tilde{v}:S^2 \to G/T$ where $T$ is a maximal torus. This
would appear to be closely related to the idea of regular deformation
of a general element $\mu$, but it says nothing about the behaviour of
the metrics.

\vspace{12pt}
We have now filled in the details of our account in the Introduction
up to equation \eqref{5.10.12}, though we have not yet shown that
$\cK$ has the structure claimed in \eqref{4.10.12}. We turn to that in
the next section.

\section{Magnetic and  holomorphic charges}
      \label{sec:charges}
We will now  show how to calculate explicitly the {\em magnetic
charges} of a monopole
which determine the homotopy class $m$ and the {\em holomorphic charges} which
determine the strata. We  will also make some conjectures about the
possible values these  can take.

In this section, $\mu$ and $k$ are as before. In addition, $T$ is a
maximal torus whose Lie algebra $\ft$ contains both $\mu$ and
$k$. Recall that a choice of Weyl chamber $C$ in $\ft$ gives rise to a
set of simple roots $\alpha_1,\ldots, \alpha_r$ and the corresponding
fundamental weights $\lambda_1,\ldots,\lambda_r$ defined by
\begin{equation}
\label{eq:fund}
2\frac{\< \alpha_i, \lambda_j\>}{\< \alpha_i, \alpha_i\>} = \delta_{ij}.
\end{equation}

We can always choose a fundamental Weyl chamber $C$ satisfying
\begin{equation}
\label{eq:simple}
\alpha_1(\mu) > 0, \dots, \alpha_s(\mu) > 0,\mbox{ and }
\alpha_{s+1}(\mu) = 0, \dots, \alpha_r(\mu) = 0,
\end{equation}
because this is just the condition that $\mu$ is in the closure
of $C$ and a particular ordering of the simple roots.

We would like to apply the corresponding fundamental weights to $k$  but this
is not possible as we only know that $k$ is in the Lie algebra of the
centraliser of $\mu$.   We can conjugate $k$ by $C(\mu)$
until it is inside the torus but then we find that $\ft \cap C(\mu) k$
is not a single point but an orbit under $\cW_\mu$ the subgroup of the
Weyl group stabilising $\mu$.  Our first result  resolves this
problem by showing that  we can pick out a unique element
$\tilde k$ of $\ft \cap C(\mu) k$.

\begin{proposition} Suppose that the moduli space $\cM(u,\mu,k)$ is
non-empty and we have fixed a maximal torus containing $\mu$,  a
fundamental Weyl chamber $C$ with $\mu$ in its closure and have
ordered the simple roots so they satisfy \eqref{eq:simple}. Then
there exists a  uniquely determined $\tk\in \ft \cap C(\mu) k$, such that
$$
\alpha_{s+1}(\tilde k) \leq  0, \dots, \alpha_r(\tilde k) \leq 0.
$$
Moreover, we have $\lambda_j(\tk) \geq 0$ for $j=1,\ldots, r$.
\label{lieprop}
\end{proposition}

We shall give the proof of this proposition in a moment. For now, we
shall use it to define the {\em charges} of the monopole to be the
non-negative integers
$$
\lambda_1(\tilde k), \dots,\lambda_r(\tilde k).
$$
They are naturally divided into {\em magnetic charges}
$$
m_1 = \lambda_1(\tilde k), \dots, m_s = \lambda_s(\tilde k)
$$
and the {\em holomorphic charges}:
$$
h_1 = \lambda_{s+1}(\tilde k), \dots, h_{r-s}  = \lambda_r(\tilde k).
$$
In some examples the simple roots have a natural ordering and it is
convenient not to re-order them. In that
case we just choose $\tilde k$ to be the unique  $\tilde k \in \ft
\cap C(\mu) k$ such that
whenever $\alpha_i(\mu) = 0$ we have $\alpha_i(\tilde k) \leq 0$.  We
then say that
$\lambda_i(\tilde k)$ is a magnetic charge if $\alpha_i(\mu) > 0$
and a holomorphic charge if $\alpha_i(\mu) = 0$.

The most important point to be made here is that it is easy to show that
$\pi_2(O_\mu) = \ZZ^s$ and the magnetic charges
determine the
homotopy class of $\phi$ the Higgs field at infinity (see for example
\cite{BerGelGel}).
The magnetic charges therefore
cannot change under
continuous deformation of a monopole.  By contrast, the holomorphic
charges can jump under continuous deformation of the monopole.

Note that the strata in the moduli space are all those monopoles with
the same $\tilde k$.

As well as being non-negative the holomorphic charges satisfy the additional
constraint that $\alpha_i(\tilde k) \leq 0$ for  all $i=s+1, \dots, r$.
This is equivalent to:
\begin{equation}
\label{eq:constraint}
\sum_{l=1}^{r-s} \frac{2\<\alpha_i, \alpha_{l+s} \>}{\<\alpha_{l+s},
\alpha_{l+s} \>}
    h_{l} +  \sum_{j=1}^s \frac{2\<\alpha_i, \alpha_j \>}{\<\alpha_j,
\alpha_j \>} m_j  \leq 0 ,
\quad\text{for}\quad i=s+1, \dots, r.
    \end{equation}

We conjecture:

\begin{conjecture}  For a given $\mu$ there are monopoles with any
collection of non-negative magnetic charges $(m_1, \dots, m_s)$.
Given a choice of magnetic
charges there are monopoles with any collection of holomorphic
charges $(h_1, \dots, h_{r-s})$
satisfying \ref{eq:constraint}.
\end{conjecture}
It should be possible to prove this result using rational maps but it
has eluded us. We can prove
however:

\begin{proposition}
\label{prop:finite} For a given $\mu$ and choice of magnetic charges there are
at most a finite number of possible holomorphic charges satisfying
\ref{eq:constraint}.
\end{proposition}
We defer the proof to the next section but note that this gives
\begin{corollary}
There are only a finite number of strata and in particular there must
be an open stratum.
\end{corollary}

\vspace{10pt}
Note that this approach gives a nice picture in terms of Dynkin diagrams. For
maximal symmetry-breaking, all charges are magnetic (i.e.\
topological) and the heuristic
is that there are $m_i$ fundamental monopoles of type $i$ for
each $i$ a node on the Dynkin diagram.
For non-maximal symmetry breaking mark each node $i$ with $\alpha_i(\mu)=0$.
Now  each Dynkin node still has associated to it the non-negative
integer $\lambda_i(\tk)$. This number is a magnetic charge $m_i$ if
$i$ is unmarked,
and again the heuristic is that there
are $m_i$ fundamental monopoles of type $i$.  If $i$ is a marked node
then $\lambda_i(\tk)$ is a holomorphic charge. This labels the strata
and can jump under continuous deformation of the monopole. The
possible holomorphic charges are constrained by
inequalities
which can be deduced from the Dynkin diagram and
\eqref{eq:constraint}.

\subsection{Proof of Proposition~\ref{lieprop}}
Let $\cW_\mu$ be the subgroup of the Weyl group
fixing $\mu$ and note that it acts transitively on the
set of all fundamental Weyl chambers with $\mu$
in their closure \cite{Hum}.

To prove first that a $\tilde k$ exists we follow Jarvis
\cite{Jar98a} and consider
the condition  $\alpha(\mu - tk') > 0$
for large enough $t$ and any $k' \in \ft \cap C(\mu)k$.    As there
are only a finite
number of roots we  can find an $\epsilon > 0$ such that for all $ t
\in (0, \epsilon]$ we have that
$\alpha(\mu - t k') = 0$ if and only if $\alpha(\mu)  = 0$ and
$\alpha(k') = 0$ and
$\alpha(\mu - t k') > 0 $ implies $\alpha(k') = 0$ and $\alpha(k') <
0$.  For any
such $t$ choose a fundamental Weyl chamber with $\mu-tk'$ in its
closure. As $t \to 0$
we see that this has $\mu$ in its closure as well.  If this is not
the fundamental
Weyl chamber we first thought of we can move it by $\sigma \in
\cW_\mu$ until it
is and then
let $\tilde k =\sigma(k')$.  Then $\mu - t\tilde k$ is in the closure of our
fundamental Weyl chamber so that $\alpha_i(\mu) > 0$ for $i=1, \dots, s$ and
$\alpha_j(\mu) = 0$ and $\alpha_j(\tilde k) \leq 0$ for $j=s+1, \dots, r$.

We will see in a moment that $\tilde k$ is unique but for now
we show that  $\lambda_i(\tilde k) \geq 0$ for all $i=1, \dots, r$.

    Consideration of the twistor construction for monopoles
shows that $\phi$ and $f$ satisfy the following non-negativity
constraint for any direction $u$.  Choose any
maximal torus $T$ so that $\phi(u),f(u)\in \ft$. Choose a
fundamental Weyl chamber whose closure contains $\phi(u)$ and
let $\alpha_1, \dots, \alpha_r$ be the corresponding
simple roots. Define the    fundamental weights $\lambda_1, \dots,
\lambda_r$  by \eqref{eq:fund} then:
$$
\lambda_i(f(z)) \geq 0\mbox{  for all }i=1, \dots, r
$$
independent of all the choices made. Note that $\tilde k$ is a
conjugate of $k$ under an element of $C(\mu)$ and hence
corresponds to the $k$ for some different monopole which
also satisfies the positivity constraint.  Hence we
must have $\lambda_i(\tilde k) \geq 0$ for all $i=1, \dots, r$.

Consider lastly the uniqueness of $\tilde k$.  So assume we have
$\tilde k$ and $\sigma(\tilde k)$
for $\sigma \in \cW_\phi$ and $\alpha_j(\tilde k) \leq 0$ and
$\alpha_j(\sigma(\tilde k)) \leq 0$
for every $i = s+1, \dots, r$.  Let $V$ be the span of the roots
$\alpha_{s+1} , \dots, \alpha_r$.
This is a root system with Weyl group $\cW_\mu$. Let $C_{ij}$ be the
inverse of the matrix
$D_{ij} = \< \alpha_i, \alpha_j\>$. Then both $C$ and $D$ are symmetric. Define
$$
\chi \colon \ft \to V
$$
by
$$
\chi(h) = \sum_{j,k=s+1}^r \alpha_j(h)C_{jk}\alpha_k.
$$
Let $\sigma_l$ be a simple root reflection for $s+1 \leq l \leq r$. Then
\begin{align*}
\chi(\sigma_l(h)) &= \sum_{j,k=s+1}^r \sigma_l(\alpha_j)(h)C_{jk} \alpha_k\\
        &=\chi(h) - \sum_{j,k=s+1}^r \frac{2\<\alpha_j,
\alpha_l\>}{\<\alpha_l, \alpha_l\>}C_{jk}  \alpha_k(h) \\
        &= \chi(h) - \frac{2 \alpha_l(h) }{\<\alpha_l, \alpha_l\>}\alpha_l.
\end{align*}
Moreover
\begin{align*}
\sigma_l(\chi(h)) &= \chi(h) - \frac{2\<\chi(h),
\alpha_l\>}{\<\alpha_l, \alpha_l\>}\\
           &=\chi(h) - \sum_{j,k=s+1}^r \alpha_j(h)C_{jk}
\frac{2\<\alpha_k, \alpha_l\>}{\<\alpha_l, \alpha_l\>}\\
          &= \chi(h) - \frac{2 \alpha_l(h) }{\<\alpha_l, \alpha_l\>}\alpha_l\\
&=\chi(\sigma_l(h)).
         \end{align*}
It follows that if $\sigma \in \cW_\mu$ then $\chi(\sigma(\tilde k)) =
\sigma(\chi(\tilde k))$. We also have
$\<\alpha_l, \chi(h)\> = \alpha_l(h)$ so that $\chi(\tilde k)$ and
$\chi(\sigma(\tilde k))$ are in the closure
of the same Weyl chamber in $V$.
Applying
Humphreys' 10.3 Lemma B \cite{Hum}  we see that $\chi(\sigma(\tilde k)) =
\sigma(\chi(\tilde k)) = \chi(\tilde k)$ and hence $\alpha_i(\tilde k
- \sigma(\tilde k)) = 0$
for $i = s+1, \dots, r$.  We have previously seen that $\lambda_i(\tilde k -
\sigma(\tilde k)) = 0$ for $i=1, \dots, s$.
Moreover the span of the $\lambda_1, \dots, \lambda_s$ is orthogonal
to the span of the
$\alpha_{s+1}, \dots, \alpha_r$ so together they must span $\ft^*$ and
hence $\tilde k = \sigma(\tilde k)$.

\subsection{Proof of Proposition~\ref{prop:finite}}
Let $\epsilon$ be the sum of all the positive roots which
are in the span of the simple roots $\alpha_{s+1}, \dots, \alpha_r$.
Notice that
$\epsilon(\tilde k) \leq 0$. Recall \cite{Hum} that a simple root reflection
$\sigma_i$ permutes all the positive roots except $\alpha_i$ which it sends to
$-\alpha_i$. So if $s+1 \leq i \leq r$ we have $\sigma_i(\epsilon) =
\epsilon - 2 \alpha_i$ so that
$$
2 \frac{\< \epsilon , \alpha_i \> }{\< \alpha_i, \alpha_i \>} = 2.
$$
So we have
\begin{align*}
\epsilon &= \sum_{j=1}^r  2 \frac{\< \epsilon , \alpha_i \> }{\<
\alpha_i, \alpha_i \>} \lambda_i\\
         &= \sum_{j=1}^s  2 \frac{\< \epsilon , \alpha_i \> }{\<
\alpha_i, \alpha_i \>} \lambda_i + \sum_{i=s+1}^r 2 \lambda_i\\
&= \sum_{j=1}^s -p_j \lambda_j + \sum_{i=s+1}^r 2 \lambda_i
\end{align*}
where $p_j \geq 0$ because if $1 \leq j \leq s$ we have $\< \epsilon,
\alpha_j\> \leq 0$.
Applying $\epsilon $ to $\tilde k$ gives
$$
0 \leq \sum_{i=1}^{r-s} h_i \leq \sum_{j=1}^s p_j m_j
$$
and as each $h_j$ is non-negative this means there can only be a
finite number of
possibilities.

\section{Examples}
\label{sec:examples}

Let $G = SU(N)$ and
$\mu$ be a diagonal matrix with eigenvalues $i\mu_1, i\mu_2 , \dots
i\mu_q$ with
multiplicities $n_1, \dots, n_q$ and assume that $\mu_1 > \mu_2 >
\dots > \mu_q$.
Choose the usual fundamental Weyl chamber. That is,
if $d$ is any diagonal matrix with entries $id_1, \dots, id_N$ then it
is in the fundamental Weyl chamber if $d_1 > d_2 > \dots > d_N$. Clearly
this has $\mu$ in its closure. Define
$x_j(d) = d_j$. Then the simple roots are $\alpha_i = x_{i+1}-x_i$ for
$i=1, \dots, N-1$.  The fundamental weights satisfy
$$
\lambda_j(d) = d_1 + \dots + d_j
$$
and a weight is magnetic if $j=n_1, n_2, \dots, n_{q-1}$ and
holomorphic otherwise.

Let $\CC^N = \CC^{n_1} \oplus \dots \oplus \CC^{n_q}$ be the corresponding
eigenvalue decomposition of $\CC^N$. Assume that on $\CC^{n_j}$ the
eigenvalues of $k$ are
$$
k_{n_1+ \dots + n_{j-1}+1} \leq k_{n_1+ \dots + n_{j-1}+2} \leq \dots \leq
k_{n_1+ \dots + n_{j}}
$$
Then $\tilde k$ is the diagonal matrix with entries $ik_1, \dots,
ik_N$.

Let $\cM_j$ be the stratum containing $\cM(u,\mu,k)$. It was shown in
\cite{Mur} that
$$
\dim\cM_j = 4 \sum_{i=1}^N (k_1 + \dots + k_i)  + \dim C(\mu)k.
$$
and hence from the definition of the strata in the Introduction,
$$
\dim(\cM(u, \mu, k)) = 4 \sum_{i=1}^N (k_1 + \dots + k_i).
$$
so the dimension is divisible by four as required for a hyperK\"ahler
manifold. In Proposition \ref{prop:fourdim} we shall show that this
result is always true.

Notice that we could find a deformation $\mu_t$ of $\mu$ by choosing $\mu_t$ to
be diagonal with entries
$i\mu_j(t)$ such that
$$
\mu_1(t) > \mu_2(t) >\mu_3(t) >\dots > \mu_N(t)
$$
and, of course, with $\mu(0) = \mu$.
It follows from known results on the moduli spaces \cite{Wei, HurMur}
that $\dim \cM(u, \mu_t,k) = \dim \cM(u, \mu,k) $. In fact the
method used in \cite{Mur} to calculate the dimension formula shows that
$ \cM(u, \mu(t), k))$ and $\cM(u, \mu,k)$ are diffeomorphic
spaces of holomorphic maps. This result was generalized to arbitrary
$G$ by Jarvis \cite{Jar98a}.

\section{Dimensions}
\label{sec:dim}

In this section we compute the dimension of the moduli space $\cM(u, \mu, m) $
by computing the dimension of $\tR(O_\mu,  m)$. We shall also compute
the dimensions of the strata and the moduli space $\cM(u,\mu,k)$ of
monopoles of type $(\mu,k)$, by computing the
dimension of $\tR(O_{\mu k},m)$.

    Fix a
maximal torus $T$, a fundamental Weyl chamber  and a set of simple
roots $\alpha_1, \dots,
\alpha_r$. For a root $\alpha$
let $\fg_\alpha$ be the $\alpha$ root space. Denote by $B$ the standard
Borel determined by this choice of simple roots. That is the Lie
algebra of $B$ contains  the root space of every simple root.
The parabolic $P$ is determined by the fact that its Lie algebra $\fp$
contains the root spaces for the negative roots $\alpha_{s+1},
\dots, \alpha_r$.

If $f\colon S^2 \to G^c/P$ is a holomorphic map then we
can use it to pull back the tangent bundle to $G^c/P$ and the Riemann-Roch
theorem tell us that
\begin{multline*}
\dim(H^0(S^2, f^{-1}(TG^c/P)) - \dim(H^1(S^2, f^{-1}(TG^c/P))\\
\quad\quad\quad = \dim(G^c/P)+  c_1(\det(f^{-1}TG^c/P))
\end{multline*}
where $\det(TG^c/P)$ is the determinant line bundle of $f^{-1}TG^c/P$
and $c_1$ denotes the first  Chern class. Because the group $G$ acts
holomorphically on $G^c/P$
every element of $\fg$ defines a holomorphic vector field on $G^c/P$
so we have a surjection of holomorphic vector bundles over $S^2$
$$
\fg \times S^2 \to f^{-1}TG^c/P \to 0
$$
and it follows from the short exact sequence in cohomology that
$$
\dim(H^1(S^2, f^{-1}(TG^c/P)) = 0.
$$
The tangent space to $\cR(G^c/P, m)$ at the function $f$ is just the subset of
sections in $H^0(S^2,  f^{-1}(TG^c/P))$ which vanish at the base point,
say
$P \in G^c/P$. This has {\it real} dimension
\begin{align*}
\dim\cR(G^c/P, m) =& 2 (\dim(H^0(S^2, f^{-1}(TG^c/P)) -\dim G^c/P)\\
            =& 2 c_1(\det(f^{-1}TG^c/P)).\\
\end{align*}

Each of the fundamental weights $\lambda_1, \dots, \lambda_s$
extend to one-dimensional representations of $P$ and
hence define homogeneous line bundles $L(\lambda_i)$
over $G^c/P$. The magnetic charges of a holomorphic
map $f$ are $m_i = - c_1(f^{-1}(L(\lambda_i)))$. Choose $\tilde k$ so that
$m_i = \lambda_i(\tilde k)$.  Then $c_1(f^{-1}(L(-\lambda))) =
\lambda(\tilde k)$ for any weight $\lambda$.

Let  $\epsilon$ be the weight defined by the
adjoint representation of $P$ on $\fp$. Then the
weight defined by the adjoint representation of $P$ on $\fg/\fp$ is
$-\epsilon$.
The bundle
$\det(T(G^c/P)$ is  then a homogeneous bundle over $G^c/P$ induced
by the character $-\epsilon$ so that
that
$$
c_1(f^{-1}(\det(TG^c/P))) = c_1(f^{-1}(L(-\epsilon))) = \epsilon(\tilde k).
$$
Hence
$$
\dim\tR(G^c/P,  m) = 2 \epsilon(\tilde k).
$$

In the case of maximal symmetry breaking where the parabolic $P$ is a Borel $B$
$$
\epsilon = \sum_{\alpha > 0} \alpha
= 2\sum_{i=1}^r \lambda_i
$$
so that
$$
\dim\tR(G/B,  m) = 4 \sum_{i=1}^r m_i.
$$

In the non-maximal symmetry breaking case we can
proceed further.  Because $\epsilon$ is a weight we know that
$\epsilon = \sum_{i=1}^r -n_i \lambda_i$
for some integers $n_i$.  We also know that $\epsilon$ is a character of $P$
so invariant under the simple root reflections $\sigma_i$ for $i=s+1,
\dots, r$.
But $\sigma_j(\epsilon) = \epsilon + n_j \alpha_j$ so that we must have
$$
\epsilon = \sum_{i=1}^s -n_i \lambda_i
$$
and hence
$$
\dim\tR(G^c/P, m) =2 \sum_{i=1}^s n_i m_i.
$$

We can obtain some further information about the $n_i$. Firstly
we note that
$$
n_i = - 2 \frac{\<\epsilon, \alpha_i\>}{\<\alpha_i, \alpha_i\>}.
$$
Also if $\rho$ is one-half the sum of the positive roots and
$\rho_\fp$ is one-half the sum of the positive roots $\alpha$
for which $\fg_{-\alpha} \subset \fp$ then we have that $\epsilon =
-2\rho + 2\rho_\fp$
and hence
\begin{align*}
n_i  &=  2\frac{\<2\rho - 2\rho_\fp, \alpha_i\>}{\<\alpha_i, \alpha_i\>}\\
              &=  2\left(1 - 2\frac{\< \rho_\fp, \alpha_i\>}
             {\<\alpha_i, \alpha_i \>} \right) \lambda_i
             \end{align*}
using the standard fact that $\rho = \sum_{i=1}^r \lambda_i$.
Hence
$$
\dim(\tR(G^c/P, m)) =4\sum_{i=1}^s \left(1 - 2\frac{\< \rho_\fp, \alpha_i\>}
             {\<\alpha_i, \alpha_i \>} \right) m_i
             $$
     which agrees with the result in \cite{Wei}. So we have
\begin{proposition}
The dimension of the moduli space $\cM(u, \mu, m)$ is
$$
4\sum_{i=1}^s \left(1 - 2\frac{\< \rho_\fp, \alpha_i\>}
             {\<\alpha_i, \alpha_i \>} \right) m_i.
$$
\end{proposition}
Notice that while the
Lie theory guarantees $\rho$ is
a weight the same may not be true of $\rho_\fp$ and hence
     expressions such as
     $$
     2\frac{\< \rho_\fp, \alpha_i\>}
             {\<\alpha_i, \alpha_i \>}
             $$
may not be integers.  This is consistent with the fact that
     for non-maximal symmetry breaking the moduli space may not be
hyperK\"ahler for the simple reason that
      its dimension is not a multiple of four.

Next we calculate $\dim\tR(O_{\mu k},  m)$ where $P_{\mu k}$ is
the parabolic
subgroup containing all the positive roots and the negative  roots
$\alpha$ where $\alpha(\mu) =
\alpha(\tilde k) = 0$ and we let $O_{\mu k} = G^c/P_{\mu k}$.  This
is the parabolic subgroup occurring in \eqref{orbitdef}.

Then $\epsilon$ is the sum of all these roots so that
$$
\epsilon(\tilde k) =  \sum_{\alpha > 0} \alpha(\tilde k)
$$
and  we have
$$
\dim\tR(G^c/P_{\mu k},  k) = 4 \left( \sum_{i=1}^s m_i +
\sum_{j=1}^{r-s} h_j \right).
$$
Hence we deduce:
\begin{proposition}
\label{prop:fourdim}
The dimension of the moduli space $\cM(u, \mu, k)$ is
$$
4 \left( \sum_{i=1}^s m_i + \sum_{j=1}^{r-s} h_j \right).
$$
In particular it is divisible by four.
\end{proposition}
Similarly for the strata, we have
\begin{corollary}
The dimension of the stratum $\cM_j$ containing  $\cM(u, \mu, k) $ is
$$
4 \left( \sum_{i=1}^s m_i + \sum_{j=1}^{r-s} h_j \right) + \dim
C(\mu) - \dim C(\mu) \cap C(k).
$$
\end{corollary}

\subsection{Acknowledgements}
The first author thanks Michael Eastwood for helpful discussions and the
Australian Research Council for support.


\end{document}